\documentclass[12pt]{revtex4}
\usepackage{amsmath,amssymb}
\usepackage{dcolumn}
\usepackage{bm}
\usepackage{color}
\usepackage{amsthm,amscd}
\usepackage{graphicx}
\newcommand{\bib}[2]{\frac{\partial {#1}}{\partial {#2}}}
\newcommand{\bbib}[3]{\frac{\partial^2 {#1}}{\partial {#2}{\partial {#3}}}}
\newcommand{\tc}[2]{\textcolor{#1}{#2}}
\def\w{{\wedge}}

\begin{document}
\title{New covariant Lagrange formulation
for field theories}
\author{Takayoshi Ootsuka}
 \email{ootsuka@cosmos.phys.ocha.ac.jp}
 \affiliation{Physics Department, Ochanomizu University, 2-1-1 Ootsuka Bunkyo Tokyo, Japan }
\begin{abstract}
A novel approach for Lagrange formulation for field theories
is proposed in terms of Kawaguchi geometry (areal metric space). 
On the extended configuration space 
$M$ for classical field theory
composed of spacetime and field configuration space, 
one can define a geometrical structure called Kawaguchi areal metric $K$ 
from the field Lagrangian
and $(M,K)$ can be regarded as Kawaguchi manifold. 
The geometrical action functional is given by $K$ 
and the dynamics of field is determined by  
covariant Euler-Lagrange equation 
derived from the variational principle of the action.
The solution to the equation becomes a minimal hypersurface on $(M,K)$
which has the same dimension as spacetime. 
We propose that this hypersurface is what we should regard as our real spacetime manifold, 
while the usual way to understand spacetime 
is to consider it as the parameter spacetime (base manifold) of a fibre bundle.
In this way, the dynamics of field and spacetime structure is unified by Kawaguchi geometry. 
The theory has the property of 
{\it strong covariance}, 
which means that it allows one to choose 
arbitrary parameter spacetime
from the total extended configuration space. 
In other words, 
we can take more wider class of coordinate transformations such that 
mix spacetime and field variables which are forbidden to be exchanged in 
conventional formalism. 
In this aspect, we call our formalism {\it fibration-free}.
This aspect also simplifies the discussion on symmetries 
such as N\"other's theorem. 
\end{abstract}
\maketitle

\pagestyle{plain}

\section{Introduction}

In the conventional Lagrange formulation
of finite dimensional systems,
the configuration space $Q$
does not have a natural geometrical structure.
The action $S[c]$ defined by the Lagrangian depends on the 
parameterisation of the curve $c:T\subset\mathbb{R}\to Q$, 
which is also not a geometrical object.

On the other hand, there exists a natural Finsler structure 
$F(x,dx)=L\left(x^i,\frac{dx^i}{dx^0},x^0\right)|dx^0|$ 
on the extended configuration space $M=\mathbb{R}\times Q$, 
and the set $(M,F)$ becomes a Finsler manifold. 
Finsler function $F$
defines the geometrical action functional
\begin{eqnarray}
 {\cal A}[\bm{c}]=\int_{\bm{c}}F
 :=\int_{T}c^{\ast}F\left(x^\mu(\tau),\frac{dx^\mu}{d\tau}(\tau)\right)d\tau
\end{eqnarray}
for an oriented curve $\bm{c} \subset M$. 
The action is independent of parameterisation 
$c:T\subset \mathbb{R} \to M$ of the curve $\bm{c}$, 
and the Euler-Lagrange equation derived from ${\cal A}[\bm{c}]$ becomes 
a geodesic equation. 
Furthermore, once this reformulation is achieved, 
the system becomes covariant, reparameterisation invariant. 
This is equivalent to saying that the system becomes independent of 
local fibration $M \to T \subset \mathbb{R}$.
Here, the 
``time'' coordinate $x^0$ and configuration coordinates $x^i$
are on the same level, 
and coordinate transformations may consist of all these variables. 
Every Lagrange mechanics could be reformulated 
in this Finsler geometry perspective~\cite{Lanczos},  
however, 
the effectiveness of this observation has not been considered so much 
by mathematicians nor physicists. 
One example is a work by Ootsuka and Tanaka~\cite{OT2}, 
which introduces Finsler geometry in the settings 
for formulating Feynman's path integral geometrically. 

The application of Finsler geometry to the finite dimensional Lagrangian system is 
still a vast area for research, but here we will proceed further 
and consider a similar geometrical formulation for field theory. 
We propose that the underlying geometry of Lagrange formulation for field theory 
is Kawaguchi geometry, 
which is a natural extension of Finsler geometry.
We can define a Kawaguchi structure (areal space metric) 
from field Lagrangian
on extended field configuration space $M=S \times Q$
which is composed of ``spacetime'' $S$ and field configuration space $Q$.

By this new formulation, 
the geometrical structure of spacetime and 
dynamics of field will be unified, 
and the Lagrange formulation for field theory 
will be reparameterisation invariant 
and independent of local fibration
$M \to S\subset \mathbb{R}^{n+1}$.
We call this property {\it strong covariant} or {\it fibration-free}.

We will first make a short review on Kawaguchi manifold 
and then define a new Lagrange formulation for field theory using Kawaguchi geometry.  
Covariant Euler-Lagrange equation and N\"other theorem are disscussed. 
We will show that the N\"other theorem becomes extremely simple in our geometrical setting. 

The fibration-free property of the formulation enables us to use more 
general transformation consisting of both spacetime and field coordinates, 
rather than the conventional fiber bundle framework.
The spacetime $S$ in the Kawaguchi geometry picture 
is a mere parameter space picked out from the extended configuration space. 
In this viewpoint, the geometrical structure of $S$ 
is not given from the beginning as in the fibre bundle picture, 
but is something that should be derived from the original Kawaguchi manifold $(M,K)$. 
Our formulation tells us to reconsider the concept of spacetime in such a way.

\section{Kawaguchi Geometry}

Kawaguchi manifold $(M, K)$ is a set of differentiable 
$N$-dimensional manifold 
$M$ and Kawaguchi $(n+1)$-dimensional areal metric $K$ which is defined as 
a function of $x=(x^\mu)$ and $dx=(dx^{\mu_0\mu_1\cdots \mu_n})$,
obeying the following homogeneity condition:
\begin{eqnarray}
K \left(x,\lambda dx \right) 
= \lambda K\left(x,dx \right), \quad \lambda > 0. \label{homo}
\end{eqnarray}
Here we assume $N > n+1$ and  
$dx^{\mu_0\mu_1\cdots \mu_n}$ is for 
$dx^{\mu_0} \w dx^{\mu_1} \w \dots \w dx^{\mu_n}$, with 
$\mu_a=0,1,2,\cdots ,N~(a=0,1,2,\cdots ,n)$.

In general, $K$ is not exactly an $(n+1)$-form, 
but nevertheless we are inclined to call this 
a {\it Kawaguchi $(n+1)$-form}, because an $(n+1)$-form is the simplest example of 
Kawaguchi form, and when carrying out the calculation, $K$ is pulled back 
to a parameter space where it becomes a standard $(n+1)$-form. 

For any $(n+1)$-vector 
$\bm{v} \in \Lambda^{n+1} T_pM$,
$K$ gives a norm 
$K(x(p),dx(\bm{v}))$.
$K$ measures the hyperarea of an oriented $(n+1)$-dimensional submanifold 
$\bm{\sigma}$ on ${M}$.  
Taking an arbitrary parameterisation 
$\sigma=(s^0,s^1,\dots ,s^{n}):S\subset \mathbb{R}^{n+1} \to M$
for
$\bm{\sigma}$, 
the hyperarea ${\cal A}$ of 
$\bm{\sigma}$ is given by 
\begin{eqnarray}
 {\cal A}[\bm{\sigma}] = \int_{\bm{\sigma}} K\left(x,dx \right) 
 :=\int_{S} 
 \sigma^\ast K\left(x^\mu(s),\frac{\partial (x^{\mu_0},\cdots ,x^{\mu_n})}
 {\partial (s^0,\cdots,s^n)}
 \right)ds^0 \w \cdots \w ds^n.
\end{eqnarray}
${\cal A}[\bm{\sigma}]$ depends on the orientation of the 
submanifold $\bm{\sigma}$, 
but by the homogeneity condition, does not depend on the 
reparameterisation $\sigma$
such that preserves an orientation. 
So $K$ can define the geometric hyperarea of $\bm{\sigma} \subset M$.

The Kawaguchi manifold $(M,K)$ is a natural generalisation of 
Riemannian manifold. 
There is a canonical way to define the areal metric 
from $g$, where $g$ is a metric of the Riemannian manifold. 
However a general Kawaguchi metric $K$ cannot be derived from 
Riemannian structure of $M$, 
as in the case of Finsler metric.

\section{Kawaguchi-Lagrange formalism for field theory}

Here we will restrict ourselves to consider local properties of the field theory 
and leave the global properties untouched. 
Consider a field theory in $(n+1)$-dimensional spacetime, 
having $D$ dimensional fiber (field configuration) space 
which is described by Lagrangian $(n+1)$-form,
\begin{eqnarray}
 L(\varphi, \partial \varphi)~dx^0 \w dx^1 \w \cdots \w dx^n.
\end{eqnarray}

Consider a manifold, 
\begin{eqnarray}
 M=\mathbb{R}^{n+1}\times \mathbb{R}^D
 =\{(x^0,x^1,\cdots,x^n,\varphi_1,\dots ,\varphi_D)\}, \label{K-mfd}
\end{eqnarray}
then we can construct Kawaguchi $(n+1)$-form $K$ on $M$ by, 
\begin{eqnarray}
 K(x,dx)=L\left(x^{n+i},\frac{dx^{n+i}_a}{dx^{012\cdots n}}\right)
 |dx^{012\cdots n}|, \label{K-func}
 \label{K}
\end{eqnarray}
where $a=0,1,2,\dots,n$, $x^{n+i}=\varphi_i \, (i=1,2,\dots,D)$ 
and $dx^{n+i}_a$ are defined by substituting 
the $a$-th component of $dx^{012\cdots n}$ by $dx^{n+i}$, that is,
\begin{eqnarray}
 dx^{n+i}_a:=dx^{n+i}\w \left(
 \iota_{\frac{\partial}{\partial x^a}}dx^{012\cdots n}
 \right).
\end{eqnarray}
For example, 
$dx^{n+i}_1=dx^0\w d\varphi^{i} \w dx^2 \w \cdots dx^n$.
With (\ref{K-mfd}) and (\ref{K-func}) together, we may now 
recognise the Lagrange field theory as a Kawaguchi manifold $(M,K)$. 
Usually, the field Lagrangians we consider are reversible systems, 
so we inserted an absolute value on the end of the formula (\ref{K-func}). 
However, when making the calculus of variation, 
we only consider submanifolds which have constant orientation, 
so this absoluteness may be ignored in the following discussion. 

Given a Kawaguchi manifold $(M,K)$,
the geometrical action is given by, 
\begin{eqnarray}
 {\cal A}[\bm{\sigma}]=\int_{\bm{\sigma}} K(x,dx),  \label{K-action}
\end{eqnarray}
where oriented $(n+1)$-hypersurface $\bm{\sigma}$ 
represents a field configuration and dynamics.
The classical solution of the field corresponds to 
the extremal $\bm{\sigma}_c$ of the action ${\cal A}$
which is the solution 
to the variational equation, $\delta {\cal A}[\bm{\sigma}_c]=0$.

Kawaguchi structure $K$ on $M$ naturally induces a geometric structure 
on the submanifold $\bm{\sigma}_c$, as in special relativity
Minkowski metric induces a proper time for the particle trajectory.
Our spacetime is better realised by the solutional space $\bm{\sigma}_c$ 
than by the parameter space (or the base space) in the bundle formulation. 
In general, the induced structure need not be a Riemannian structure.
That leads to a new question, how to construct a unified theory 
including gravity by the Kawaguchi-Lagrange formalism.

\subsection{covariant Euler-Lagrange equation}

Now we consider the covariant equation of motion derived from the action (\ref{K-action}).
The variation of $K$ becomes
\begin{eqnarray}
 \delta K&=&\delta x^\mu 
 \left\{ 
 \frac{\partial K}{\partial x^\mu}
 -d\left(
 \frac1{p!}
 \frac{\partial K}{\partial dx^{\mu\mu_1\mu_2\cdots \mu_n}}
 \right) \w dx^{\mu_1\mu_2\cdots \mu_n}
\right\} \nonumber \\
 && \hspace{3cm}
 +d\left\{\delta x^\mu
 \left(
 \frac1{n!}
 \frac{\partial K}{\partial dx^{\mu\mu_1\mu_2\cdots \mu_n}}
 dx^{\mu_1\mu_2\cdots \mu_n}\right)
 \right\}.
\end{eqnarray}
And so 
the {\it covariant Euler-Lagrange equation} becomes, 
\begin{eqnarray}
 {\cal EL}_\mu(K)=0, \label{el-eq}
\end{eqnarray}
where the Euler-Lagrange derivation defined by 
\begin{eqnarray}
 && \hspace{-12pt}
 {\cal EL}_\mu(K)=
 \frac{\partial K}{\partial x^\mu}
 -d\left(
 \frac1{n!}
 \frac{\partial K}{\partial dx^{\mu\mu_1\mu_2\cdots \mu_n}}
 dx^{\mu_1\mu_2\cdots \mu_n} \right).
\end{eqnarray}
This is the geometrical Euler-Lagrange equation 
which is not only coordinate-free, 
but also independent of parameterisation.
Here we dropped a notation by a pullback by $\sigma$, 
namely, $\delta K = \delta (\sigma^\ast K)$, 
$dx^{\mu_1\mu_2\cdots \mu_n}=\sigma^\ast (dx^{\mu_1\mu_2\cdots \mu_n})$,
and etc.

For concrete calculation, we take the convenient
parameterisation of the submanifold 
$\sigma(s):S\subset\mathbb{R}^{n+1} \to M$ and consider a pullback by $\sigma(s)$. 
If we take conventional spacetime parameter $s^a=x^a \, (a=0,1,2\cdots,n)$, 
then the equations (\ref{el-eq}) become the usual Euler-Lagrange equations
\tc{black}{plus energy-momentum conservatation laws}. 
The freedom on the choice of spacetime parameters, covariance,
is the essential and advantageous points of our theory.

\subsection{covariant N\"other currents}

Next we will generalise N\"other theorem in our formulation.
The symmetry of the system is given by the 
vector fields on Kawaguchi manifold $(M,K)$, 
and there is no distinction between external (spacetime) 
and internal (field) symmetries. 
We define the Lie derivative ${\cal L}_{v}$ acting on the Kawaguchi form $K$ by
the first prolongation of a vector field $v$,

\begin{eqnarray}
 {\cal L}_v K:=v^\mu
 \frac{\partial K}{\partial x^\mu}
 +dv^\mu \w dx^{\mu_1\cdots\mu_n}
 \frac{\partial K}{\partial dx^{\mu \mu_1\cdots\mu_n}}.
\end{eqnarray}
We say that a vector field $v$ is a Killing vector field
of Kawaguchi form $K$, when 
\begin{eqnarray}
 {\cal L}_v K=dB,
\end{eqnarray}
where $B$ is a function of $x^\mu$ only and $dB$ is the exterior derivative of $B$. 
By this Killing vector field $v$, 
the {\it covariant N\"other current} corresponding to this symmetry is defined by 
\begin{eqnarray}
  J=v^\mu
 \left(
 \frac1{n!}
 \frac{\partial K}{\partial dx^{\mu\mu_1\cdots\mu_n}}
 dx^{\mu_1\mu_2\cdots \mu_n}\right)-B. \label{Noether}
\end{eqnarray}
This of course satisfies 
\begin{eqnarray}
 dJ=0,\label{dJ}
\end{eqnarray}
under the on-shell ${\cal EL}_\mu(K)=0$ condition. 
$(\ref{Noether})$ or $(\ref{dJ})$ are not the standard differential forms, 
but when we consider its pull back to the spacetime parameter space, 
they become the standard differential forms. 
When we use the conventional spacetime parameterisation 
$s^a=x^a~(a=0,1,2, \cdots ,n)$, 
the well-known form of N\"other current can be obtained. 

\section{Examples} \label{sec.examples}

\subsection{Nambu-Goto string}

Let $(M,\eta)$ be an $(N+1)$-dimensional Minkowski manifold, 
with a Minkowski metric $\eta$ and 
$X^I \, (I=0,1,2,\cdots ,N)$ be coordinates of $M$. 
Then we can define Kawaguchi $2$-form by
\begin{eqnarray}
 && \hspace{-12pt}
 K(X,dX)=\sqrt{-\textstyle{\frac12} dX_{IJ}dX^{IJ}}, \, dX^{IJ}=dX^I \w dX^J,
\end{eqnarray} 
which describe free string motion; the Nambu-Goto string. 
$dX_{IJ}$ are defined by $dX_{IJ}=dX_I \w dX_J$ and $dX_I=\eta_{IK}dX^K$. 
Kawaguchi 2-form $K$ is easily checked that it satisfies 
homogeneity condition $(\ref{homo})$. 
Nambu-Goto string is naturally described by Kawaguchi geometry~\cite{Ingarden}
in the same way of relativistic free particle naturally described by Finsler geometry. 
Riemannian manifold has natural Finsler or Kawaguchi structure 
and the induced structure on submanifolds of $M$ have also Riemannian structure. 
However, in the semi-Riemannian case, 
submanifolds do not have a natural Riemannian structure. 
In that case, we must consider
Zeeman structure~\cite{Zeeman} 
or generalised Busemann-Tamassy structure defined in \cite{OT2}.
 
Covariant Euler-Lagrange equation derived from the above Kawaguchi form is 
\begin{eqnarray}
  d\left(\frac{dX_{IJ}}{\sqrt{-\textstyle{\frac12}dX_{KL}dX^{KL}}}
  dX^J \right)=0.
\end{eqnarray}
Using a parameterisation $(s^0,s^1)$, the equation becomes the usual form, 
\begin{eqnarray}
  d\left\{
 \frac{\frac{\partial (X_I,X_J)}{\partial (s^0,s^1)}
 \left(\bib{X^J}{s^0}ds^0+\bib{X^J}{s^1}ds^1\right)
  }{\textstyle\sqrt{
-\frac12 \bib{(X_K,X_L)}{(s^0,s^1)}
\bib{(X^K,X^L)}{(s^0,s^1)}
}}
\right\}=0.
\end{eqnarray}
If we choose a specific parameterisation 
$\sigma(s,t):S\subset \mathbb{R}^2 \to M$; 
a Kawaguchi proper areal parameter $s$ and $t$, which satisfy 
\begin{eqnarray}
 1=\sqrt{-\frac12 \frac{\partial (X^I,X^J)}{\partial (s,t)}
 \frac{\partial(X_I,X_J)}{\partial (s, t)}}, 
\end{eqnarray}
and the other condition,
\begin{eqnarray}
 \eta\left(
 \sigma_\ast\left(\bib{}{t}\right),\sigma_\ast\left(\bib{}{t}\right)
 \right)= \frac{\partial X_\mu}{\partial t}
 \frac{\partial X^\mu}{\partial t}=0,
\end{eqnarray}
which is a natural condition in the Minkowski case, 
then the equation of motion of string becomes 
\begin{eqnarray}
 \frac{\partial^2 X^I}{\partial s \partial t}=0, \quad
 (I=0,1,2,\cdots, N).
\end{eqnarray}
This equation is the counterpart of the equation of relativistic particle using proper time.

\subsection{$(1+1)$-dimensional complex scalar field theory}

Let us consider two dimensional complex scalar field
with a Lagrangian $L=\partial \phi \cdot \partial \bar{\phi}-V(|\phi|^2)$.
We take $M=\mathbb{R}^2\times \mathbb{C}$, and the coordinates 
$(x^\mu)=(t,x,\phi,\bar{\phi})$. 
Then on $M$ we can define a Kawaguchi 2-form
\begin{eqnarray}
 K(x,dx)=\frac{dx^{12}dx^{13}-dx^{02}dx^{03}}{dx^{01}}
 -V dx^{01}.
\end{eqnarray}

The covariant Euler-Lagrange equations derived from Kawaguchi 2-form are 
\begin{eqnarray}
&&
d\left[
\biggl\{
\frac{dx^{12}dx^{13}-dx^{02}dx^{03}}{(dx^{01})^2}
+V \biggr\}dx^1 
+\frac{dx^{03}}{dx^{01}}dx^2
+\frac{dx^{02}}{dx^{01}}dx^3
\right]
=0,\\
&&
d\left[
\biggl\{
\frac{dx^{12}dx^{13}-dx^{02}dx^{03}}{(dx^{01})^2}
+V \biggr\}dx^1 
+\frac{dx^{13}}{dx^{01}}dx^2
+\frac{dx^{12}}{dx^{01}}dx^3
\right]
=0,\\
&&
 d\biggl(
 \frac{dx^{03}}{dx^{01}}dx^0
 -\frac{dx^{13}}{dx^{01}}dx^1
 \biggr)
 +V'x^3 dx^{01}=0,\\
&&
 d\biggl(
 \frac{dx^{02}}{dx^{01}} dx^0 
 -\frac{dx^{12}}{dx^{01}}dx^1
 \biggr)
 +V'x^2 dx^{01}=0.
\end{eqnarray}
The Kawaguchi 2-form $K$ is invariant by the generators, 
\begin{eqnarray}
 &&
 \hspace{-12pt}
 v_0=\frac{\partial}{\partial x^0}, \quad
 v_1=\frac{\partial}{\partial x^1}, \quad
 w=i\left(
 x^2\frac{\partial}{\partial x^2}-
 x^3 \frac{\partial}{\partial x^3}\right).
\end{eqnarray}
$v_0,\,v_1$ and $w$ are vector fields on $M$ and we can easily 
check ${\cal L}_{v_0} K={\cal L}_{v_{1}}K={\cal L}_w K=0$. 
So from the previous N\"other theorem~$(\ref{Noether})$, 
\begin{eqnarray}
 &&
 T_0=-\frac{dx^{12}dx^{13}-dx^{02}dx^{03}}{(dx^{01})^2}dx^1
     -\frac{dx^{03}}{dx^{01}}dx^2-\frac{dx^{02}}{dx^{01}}dx^3,
  \nonumber \\
 &&
 T_1=\frac{dx^{12}dx^{13}-dx^{02}dx^{03}}{(dx^{01})^2}dx^0
     +\frac{dx^{13}}{dx^{01}}dx^2+\frac{dx^{12}}{dx^{01}}dx^3,
    \nonumber\\
 &&
 J=i\left\{
 \left(x^2\frac{dx^{03}}{dx^{01}}-x^3\frac{dx^{02}}{dx^{01}}\right)dx^0 
 -\left(x^2\frac{dx^{13}}{dx^{01}}-x^3\frac{dx^{12}}{dx^{01}}\right)dx^1
 \right\},
\end{eqnarray} 
are the covariant N\"other currents of Killing vectors, respectively. 
Comparing to the conventional terminology, 
the current $(T_0,T_1)$ is 
the covariant form of energy-momentum current and 
$J$ is the covariant form of electric charge current.
In our framework, 
these conserved quantities are all derived on the same level, 
and in a simple manner.

\subsection{$(1+3)$-dimensional Maxwell theory}

Free Maxwell field can be described by
Kawaguchi manifold $(M,K)$
where 
\begin{eqnarray}
 &&M=\mathbb{R}^8=\{(x^0,x^1,x^2,x^3,A_0,A_1,A_2,A_3)\}=\{(x^\mu)\},
 \nonumber \\
 &&K=\frac{1}{2dx^{0123}}\left\{
 (dx^{5123}-dx^{0423})^2+(dx^{6123}-dx^{0143})^2 
 +(dx^{7123}-dx^{0124})^2
 \right. \nonumber \\
 && \left. \hspace{24pt}
-(dx^{0623}-dx^{0153})^2 
 -(dx^{0723}-dx^{0125})^2-(dx^{0173}-dx^{0126})^2 \right\}.
\end{eqnarray}
We define 
\begin{eqnarray}
 T_\mu=\frac1{3!}\bib{K}{dx^{\mu\nu_1\nu_2\nu_3}}dx^{\nu_1\nu_2\nu_3},
\end{eqnarray}
then, the covariant Euler-Lagrange equations can be written by
\begin{eqnarray}
 dT_{\mu}=0, \quad (\mu=0,1,2,\cdots, 7).
\end{eqnarray}
We can easily check that the following vectors 
are the Killing vectors of $K$,
\begin{eqnarray}
 \hspace{-24pt}
 &\displaystyle
 v_\mu=\frac{\partial}{\partial x^\mu}, \\
 \hspace{-12pt}
 &\displaystyle
 {\ell}_{ab}=x_a \frac{\partial}{\partial x^b}
 -x_b \frac{\partial}{\partial x^a}
 +x^{4+a} \frac{\partial}{\partial x_{4+b}}
 -x^{4+b} \frac{\partial}{\partial x_{4+a}},
\end{eqnarray}
where $\mu=0,1,2,\dots , 7$,
$a,b=0,1,2,3$, $x_a=\eta_{ac}x^c$, 
$\frac{\partial}{\partial x_{4+a}}=\eta_{ac}
\frac{\partial}{\partial x^{4+c}}$, 
and $\eta_{ab}$ is 
Minkowski metric on $\mathbb{R}^4$. 
Covariant
N\"other currents corresponding to $v_\mu$ (resp. $l_{ab}$)
are $T_\mu$ (resp. $L_{ab}$).
The exact form of these N\"other currents are shown in the appendix. 

Here we would like to comment about gauge symmetry
in our formulation. 
The Maxwell field has $U(1)$ gauge symmetry
whose Killing vector field is 
\begin{eqnarray}
 {\cal G}=\sum_{a=0}^3 \frac{\partial \Lambda}{\partial x^a}
 \frac{\partial}{\partial x^{4+a}}, 
\end{eqnarray}
where $\Lambda=\Lambda(x^0,x^1,x^2,x^3)$ 
is an arbitrary function of conventional spacetime coordinates $(x^0,x^1,x^2,x^3)$. 
This symmetry is not general enough for our formulation, 
since it is assumed that the parameter space is already chosen to be 
the conventional spacetime, 
and as a consequence $\Lambda$ cannot depend on $(x^4,x^5,x^6,x^7).$ 
However, 
there is a generalisation of the gauge symmetry which is 
suited for our covariant formulation. 
The generalised $U(1)$ gauge symmetry is given by, 
\begin{eqnarray}
 \tilde{\cal G}=\sum_{a=0}^3
 \frac{d\tilde{\Lambda} \w \iota_{\partial_a}dx^{0123}}{dx^{0123}}
 \frac{\partial}{\partial x^{4+a}}.
\end{eqnarray} 
Here $\tilde{\Lambda}$ is an arbitrary function of $x^\mu, \, \mu=0,1,\cdots ,7$. 
$\tilde{\cal G}$ is called a generalised vector field 
whose coefficients are dependent on not only $x^\mu$ but also on $dx^{\mu_0\mu_1\mu_2\mu_3}$
and their symmetry is called a generalised symmetry.
In this way, we may generalise gauge principle in our new Lagrange formalism.




\section{Hilbert form for field theory} 
Similar as in the case of Finsler-Lagrange formulation,
we can define a covariant Cartan form, Hilbert form, 
on the extended velocity phase space 
(Grassmannian bundle) 
for our Kawaguchi-Lagrange formulation for field theory. 
It corresponds to generalisation of Hilbert form to field theory 
also called the Carath\'eodory form~\cite{Saunders2003}.

Constructing Hilbert form in Finsler-Lagrange formulation 
is quite simple~\cite{Chern-Chen-Lam}, 
and in Kawaguchi-Lagrange formulation the similar procedure holds. 
From the homogeneity condition of Kawaguchi form $K$, we have 
\begin{eqnarray}
 && p_{\mu_0\mu_1\cdots \mu_n}(x,dx):=\bib{K}{dx^{\mu_0\mu_1\cdots\mu_n}}, 
  \\
 &&K(x,dx)=\frac{1}{(n+1)!}
 \, p_{\mu_0\mu_1\cdots\mu_n}(x,dx)dx^{\mu_0\mu_1\cdots \mu_n}. 
\end{eqnarray}
Substituting the canonical momenta $p_{\mu_0\mu_1\cdots \mu_n}(x,dx)$ 
by functions of coordinates of Grassmannian bundle $(x^\mu,y^{\mu_0\mu_1\dots \mu_n})$, 
we obtain 
\begin{eqnarray}
 \Theta=\frac{1}{(n+1)!}
 \, p_{\mu_0\mu_1\cdots\mu_n}(x,y)dx^{\mu_0\mu_1\cdots \mu_n}, \label{h-form}
\end{eqnarray}
on the total space. 
This is the Hilbert-Carath\'eodory $(n+1)$-form 
which is covariant or independent of the choice of the fibration.

\section{Discussions} 

We proposed in this letter, that Lagrange field theories should be regarded as
Kawaguchi geometry. 
Kawaguchi-Lagrange formulation extends the 
traditional concept of general covariance of the theory to a broader sense. 

In the conventional framework, we have to start the
theory by fixing the fibre bundle structure $M \to S$ and 
we often assume the geometric structure of base space (spacetime) $S$. 
The traditional covariance only means invariance of the solution space
by the map $f:S \to S$ or
independence of adopting base space coordinates. 

On the other hand, in the Kawaguchi perspective, 
covariance means independence on the choice of 
fibration $M \to S$. 
So in our covariant theory, 
it is possible to take free choice of spacetime $S$ 
and make coordinate transformations of wider class 
that may mix spacetime and field variables. 

Since our formulation does not distinguish between 
spacetime and internal spaces, 
discussion on symmetries such as N\"other's theorem becomes extremely simple,  
and we were able to derive energy-momentum current of general relativity 
quite easily~\cite{OTY1}.

Due to the free choice of parameterisation 
and its extended flexibility of coordinate transformation,
it is expected that our formulation 
is powerful for the discussion 
of non-perturbative effects and renormalisation theory.

Also, our Kawaguchi Lagrange formulation would 
lead us to a new covariant Hamilton formulation, 
with the use of multi-contact geometry~\cite{Ootsuka}.

Higher order extensions of this Finsler-Lagrange 
and Kawaguchi-Lagrange formulation 
can be also considered~\cite{OTY2}. 

It is worthwhile to emphasise that against conventional viewpoint, 
our formulation tells us that the dynamics of fields determine the 
geometrical structure of the spacetime. 
Therefore, reconsidering physical theories such as fields on curved spacetime, 
or general relativity in this way should be an interesting and promising problem.

Although being a very important example, spinor field is not treated in our letter. 
In Finsler-Kawaguchi context, 
we still do not know how the spinor structure could be introduced 
in the theory as in the same way as Maxwell field.

As Lagrange mechanics should be formulated on Finsler manifold, 
the field theory should be formulated on Kawaguchi manifold 
from the very begining rather than starting with the conventional Lagrangian.
This perspective allows one to consider symmetry and conservation laws more clearly, 
moreover, it has the potential to describe irreversible systems 
or systems with hysteresis behaviour, 
which still has no sufficient mathematical description.

Finally, the formulation of the Feynman's path integral for field theory 
using this geometrical insight may be possible, 
following the similar way of~\cite{OT2}. 
We expect a new fibration free perspective applied to quantum field theory 
will give us a more profound understanding of problems 
such as anomaly and quantisation of gauge fields. 

\section{acknowledgments}
The work is greatly inspired by late Yasutaka Suzuki.                      
We thank Reiji Sugano, Erico Tanaka, Ryoko Yahagi, Masahiro Morikawa,
Lajos Tamassy and Laszlo Kozma
for creative discussions. 
This work was supported by the JSPS Institutional Program for Young Researcher
Overseas Visits.

\section{Notation and Appendix}

\subsubsection{simple notation}

Our notation used in the text may look more complicated 
than the conventional Lagrange formalulation. 
However, that was for easy understanding  and for actual calculation. 
There exists a simple notation for discussion of general theory,
which we present here. 
Let $(M,K)$ be a Kawaguchi manifold, with ${\rm dim}M=N$, 
then for indices $\mu=0,1,\cdots N$, 
we define $\mu'$ by 
\begin{eqnarray}
 \mu':=\mbox{(set of $n$ indices which does not include $\mu$)}.
\end{eqnarray}
Using this notation, 
the covariant Euler-Lagrange equation and N\"other current becomes,
\begin{eqnarray}
 &&{\cal EL}_\mu(K)=
  \frac{\partial K}{\partial x^\mu}
 -d\left(\bib{K}{dx^{\mu\mu'}}dx^{\mu'}\right), \\
 &&J=v^\mu \bib{K}{dx^{\mu\mu'}}dx^{\mu'},
\end{eqnarray}
as in the case of $2$-dimensional field theory.  
Here, $\mu'$ runs all the $n$ indices excluding $\mu$, 
so we define the contraction over $\mu'$ as 
\begin{eqnarray}
 A_{\mu\mu'}B^{\mu'}:=\frac{1}{n!}A_{\mu\nu_1\nu_2\cdots \nu_n}
 B^{\nu_1\nu_2\cdots \nu_n}, \, (\nu_i\neq \mu).
\end{eqnarray}

\subsubsection{higher derivative}

If we expand $d\left(\bib{K}{dx^{\mu\mu'}}dx^{\mu'}\right)$
in the $(\ref{el-eq})$, it becomes
\begin{eqnarray}
 \bbib{K}{x^\nu}{dx^{\mu\mu'}}dx^{\nu\mu'}
 +\bbib{K}{dx^{\nu\nu'}}{dx^{\mu\mu'}}d^2x^{\nu\nu'}\wedge dx^{\mu'}.
\end{eqnarray}
This is proved by using arbitrary paramaterisation 
$\sigma:S \subset \mathbb{R}^{n+1} \to M$ and calcuration of 
$\sigma^{\ast} d\left(\bib{K}{dx^{\mu\mu'}}dx^{\mu'}\right)$.
And here the pull-back of $d^2x^{\nu\nu'}\wedge dx^{\mu'}$ is
\begin{eqnarray}
 && \hspace{-12pt}
 \sigma^\ast \left(d^2x^{\nu\nu'}\wedge dx^{\mu'}\right)
 =\frac{\partial \left(
 \frac{\partial (x^\nu,x^{\nu'})}{\partial (s^0,s^{0'})},x^{\mu'}
 \right)}{\partial (s^0,s^{0'})} ds^0 \wedge ds^{0'}.
\end{eqnarray}
We can also consider $d^2x^{\nu\nu'}\wedge dx^{\mu'}$ defined above.
This corresponds to the 2nd derivative of $dx^{\mu\mu'}$ or
acceleration for field theory.

\subsubsection{remarks about formulae}

Though we omitted $\sigma^\ast$ in our formulae for convenience, 
and also supported by reparameterisation invariance,
we still should not forget that our formulae are pull-backed formulae.
For example, 
\begin{eqnarray}
 dx^{[\mu\mu'}dx^{\nu]\nu'}=0, \quad
 d^2 x^{[\mu\mu'}\wedge dx^{\nu]}=0,
\end{eqnarray}
are proved where $[\quad]$ means anti-symmetrisation.
Since these formulae in our text can be expressed differently 
by using the above conditions, they have formal ambiguities before 
the pull back.
Nevertheless the pull backed formulae are always unique 
on the parameter space.

\subsubsection{Covariant N\"other currents of Maxwell field}

We will show some covariant currents of the Maxwell field.
If we pull-back the following currents by
the conventional parameterisation $s^a=x^a \, (a=0,1,2,3)$,
then they are equal to conventional current 3-forms.
\begin{eqnarray}
 && \hspace{-36pt}
 T_0=\frac12
 \left\{
  -\left(\frac{dx^{5123}-dx^{0423}}{dx^{0123}}\right)^2
  -\left(\frac{dx^{6123}-dx^{0143}}{dx^{0123}}\right)^2
  -\left(\frac{dx^{7123}-dx^{0124}}{dx^{0123}}\right)^2
 \right. \nonumber \\
 &&
 \left.
  +\left(\frac{dx^{0623}-dx^{0153}}{dx^{0123}}\right)^2
  +\left(\frac{dx^{0723}-dx^{0125}}{dx^{0123}}\right)^2
  +\left(\frac{dx^{0173}-dx^{0126}}{dx^{0123}}\right)^2
 \right\}dx^{123}
 \nonumber \\
 &&
 +\frac{dx^{0623}-dx^{0153}}{dx^{0123}}(dx^{623}-dx^{153})
 +\frac{dx^{5123}-dx^{0423}}{dx^{0123}}dx^{423}
 \nonumber \\
 &&
 +\frac{dx^{0723}-dx^{0125}}{dx^{0123}}(dx^{723}-dx^{125})
 +\frac{dx^{6123}-dx^{0143}}{dx^{0123}}dx^{143}
 \nonumber \\
 &&
 +\frac{dx^{0173}-dx^{0126}}{dx^{0123}}(dx^{173}-dx^{126})
 +\frac{dx^{7123}-dx^{0124}}{dx^{0123}}dx^{124},
 \\
 && \hspace{-36pt}
 T_4=-\left(\frac{dx^{5123}-dx^{0423}}{dx^{0123}}\right)dx^{023}
 +\left(\frac{dx^{6123}-dx^{0143}}{dx^{0123}}\right)dx^{013}
 -\left(\frac{dx^{7123}-dx^{0124}}{dx^{0123}}\right)dx^{012}, \\
 && \hspace{-36pt}
 T_5=-\left(\frac{dx^{5123}-dx^{0423}}{dx^{0123}}\right)dx^{123}
 +\left(\frac{dx^{0623}-dx^{0153}}{dx^{0123}}\right)dx^{013}
 -\left(\frac{dx^{0723}-dx^{0125}}{dx^{0123}}\right)dx^{012},
 \\
 && \hspace{-36pt}
 L_{12}= x^1 \left[ 
 \left\{
 -\left( \frac{dx^{5123}-dx^{0423}}{dx^{0123}}\right)^2
 -\left( \frac{dx^{6123}-dx^{0143}}{dx^{0123}}\right)^2
 -\left( \frac{dx^{7123}-dx^{0124}}{dx^{0123}}\right)^2
 \right.
 \right.
 \nonumber \\
 &&
 \left.
 +\left( \frac{dx^{0623}-dx^{0153}}{dx^{0123}}\right)^2
 +\left( \frac{dx^{0723}-dx^{0125}}{dx^{0123}}\right)^2
 +\left( \frac{dx^{0173}-dx^{0126}}{dx^{0123}}\right)^2
 \right\}dx^{013}
 \nonumber \\
 &&
 -\frac{dx^{5123}-dx^{0423}}{dx^{0123}}
 \left( dx^{513} - dx^{043} \right)
 -\frac{dx^{6123}-dx^{0143}}{dx^{0123}}dx^{613}
 -\frac{dx^{7123}-dx^{0124}}{dx^{0123}}
 \left( dx^{713}- dx^{014}\right)
 \nonumber \\
 &&
 +\frac{dx^{0623}-dx^{0153}}{dx^{0123}}dx^{063}
 +\frac{dx^{0723}-dx^{0125}}{dx^{0123}}
 \left( dx^{073}- dx^{015} \right)
 -\frac{dx^{0173}-dx^{0126}}{dx^{0123}}dx^{016}
 \Biggr]
 \nonumber \\
 &&
 -x^2
 \left[
 \left\{
 -
 \left(
 \frac{dx^{5123}-dx^{0423}}{dx^{0123}}
 \right)^2
 -\left(
 \frac{dx^{6123}-dx^{0143}}{dx^{0123}}
 \right)^2
 -\left(
 \frac{dx^{7123}-dx^{0124}}{dx^{0123}}
 \right)^2
 \right.\right. \nonumber \\
&&
 \left. 
 +\left(
 \frac{dx^{0623}-dx^{0153}}{dx^{0123}}
 \right)^2
 +\left(
 \frac{dx^{0723}-dx^{0125}}{dx^{0123}}
 \right)^2
 +\left(
 \frac{dx^{0173}-dx^{0126}}{dx^{0123}}
 \right)^2
 \right\} dx^{023} \nonumber
\\
&&
 +\frac{dx^{5123}-dx^{0423}}{dx^{0123}}dx^{523}
 +\frac{dx^{6123}-dx^{0143}}{dx^{0123}}
 \left(dx^{623}-dx^{043}\right)
 +\frac{dx^{7123}-dx^{0124}}{dx^{0123}}
 \left(dx^{723}-dx^{024} \right)
\nonumber
\\
&&
 +\frac{dx^{0623}-dx^{0153}}{dx^{0123}}dx^{053}
 +\frac{dx^{0723}-dx^{0125}}{dx^{0123}}dx^{025}
 -\frac{dx^{0173}-dx^{0126}}{dx^{0123}}
 \left(dx^{073}-dx^{026} \right)
 \Biggr] \nonumber
 \\
 &&
 +x^5\left(
 -\frac{dx^{6123}-dx^{0143}}{dx^{0123}}dx^{123}
 -\frac{dx^{0623}-dx^{0153}}{dx^{0123}}dx^{023}
 +\frac{dx^{0173}-dx^{0126}}{dx^{0123}}dx^{012}
 \right)
 \nonumber \\
 &&
 -x^6\left(
 -\frac{dx^{5123}-dx^{0423}}{dx^{0123}}dx^{123}
 -\frac{dx^{0623}-dx^{0153}}{dx^{0123}}dx^{013}
 +\frac{dx^{0723}-dx^{0125}}{dx^{0123}}dx^{012}
 \right)
\end{eqnarray}

\bibliography{reference2011}

\end{document}